\documentstyle[aps,preprint]{revtex}

\begin{document}
\draft
\title{Failure of Fermi-liquid theory at strong coupling}

\author{V. Jani\v{s}}
\address{Institute of Physics, Academy of Sciences of the Czech
Republic,\\
CZ-18040 Praha 8, Czech Republic}
\maketitle
\begin{abstract}
We investigate quantitatively the effects of strong electron-electron
coupling onto the dynamics of lattice electrons. To that purpose the
self-consistent version of the bubble-chain approximation at zero
temperature and half filling of the Anderson (Hubbard) model is used.
Special attention is paid to a critical region of an electron-hole
correlation function shaping the transition from weak to strong interaction.
We find an analytic solution with Fermi-liquid properties on the
weak-coupling side of the critical region around the two-particle pole. It
is shown that Fermi-liquid theory does not lead to a consistent behavior
of the self-consistent solution on the strong-coupling side of the
critical region.
\end{abstract}
\vspace{2cm}
PACS Numbers: 71.27+a, 71.28+d

\newpage

The Anderson and Hubbard models provide a microscopic description of the
effects of electron-electron correlations onto the dynamics of a lattice
electron gas. Especially recently the single-impurity Anderson model (SIAM)\
has newly attracted attention of theorists because of its role in the exact
description of the Hubbard model in $d=\infty $ \cite{gk92,vjdv92}.
However, the two models behave at strong coupling in qualitatively different
manner. At half filling and zero temperature, we expect
a Kondo-like behavior, i.e. a narrow resonance at the Fermi level,
in the SIAM, while the
Hubbard model in $d=\infty $, when the antiferromagnetic LRO is suppressed,
turns insulating. To understand the differences in behavior of these two
related models, it is necessary to have an approximation reliable at
intermediate and strong coupling for both Anderson as well as for the
Hubbard model.

Although we know much about the SIAM\ from the Bethe-ansatz solution \cite
{and85} this method has yet proved inefficient in the Hubbard model in $%
d=\infty $ in spite of an exact transformation of the $d=\infty$ Hubbard
model onto a SIAM with a
self-consistent condition. The only technique equally well applicable to the
SIAM and to the lattice models is  many-body perturbation theory
summed via Feynman diagrams.

We know from earlier studies on the SIAM \cite{su67,ls68}
that only self-consistent (renormalized) sums of diagrams can provide
reliable approximations at intermediate and strong coupling.
Otherwise we cannot evade an unphysical RPA pole in a two-particle
Green function \cite{jrsdcm65}.

Renormalized sums of Feynman diagrams for the Hubbard model in $d=\infty $ at
weak coupling
were studied recently \cite{mm91,js95}. It was shown \cite{js95} that
self-consistent, renormalized sums of the
RPA-type can be used at any temperature and in principle at weak
as well as at strong coupling. However, it is not straightforward to
extrapolate such theories consistently to the strong-coupling regime. There
is no analytic solution to these advanced renormalized sums and
numerical solutions break down before the strong-coupling limit is reached.
The numerical troubles arise when we are approaching the RPA pole (singularity)
in an electron-hole correlation function.
There is then no effective way to make the iterations converge in the
strong-coupling regime. It is then crucial to decide from analytic estimates
how the pole in
the two-particle function is approached by the full, numerically unreachable
solution.

The transition from the weak to the
strong coupling regime can hence be investigated only analytically using some
assumptions. First such an analytic study in the SIAM was done in ref. \cite
{ha69}, where a low-frequency approximation was used to estimate the behavior
of Suhl's renormalized RPA in the critical region of the two-particle pole.
The aim of this paper is to reinvestigate the transition region between weak
and strong coupling regimes in the SIAM and the Hubbard model in $d=\infty$.
We extend Hamann's approach from ref.~\cite{ha69}  based on a Fermi-liquid,
low-frequency expansion for electron-hole bubbles and show how dominant
contributions to the
self-consistent solution can analytically be estimated when the two-particle
pole is being approached. As an example we use the self-consistent
version of the bubble-chain (shielded interaction) approximation
\cite{mm91,js95} applied to the SIAM at half filling and zero temperature.
This approximation, in contrast to earlier theories \cite{su67,ls68,ha69},
represents a thermodynamically consistent and conserving theory \cite{js95}
applicable to the SIAM as well as to the Hubbard model. As a result we obtain
a set of algebraic equations the solution of which forms a
Fermi-liquid at weak coupling, but leads to inconsistent results
at intermediate and strong couplings. This inconsistency is explained by the
failure of the Fermi-liquid, low-frequency ansatz to capture all the
relevant features of the full solution at intermediate and strong coupling.
It is necessary to take into account also incoherent, non-Fermi-liquid
contributions to the two-particle Green function to reach a consistent
behavior of diagrammatic approximations at intermediate and strong coupling.

The bubble-chain approximation for the self-energy can generally be written
in the spin-polarized version as \cite{mm91,js95}
\begin{eqnarray}
\label{eq1}\Sigma _\sigma (i\omega _n)&=&-\frac{U^2}{2\beta }\sum_{m=-\infty
}^\infty G_\sigma (i\omega _n+i\nu _m) \frac{X_{-\sigma }(i\nu _m)}{%
1-U^2X_{\uparrow }(i\nu _m)X_{\downarrow }(i\nu _m)}\, ,
\end{eqnarray}
where $X_\sigma (i\nu _m)$ is a contribution due to the electron-hole bubble,
\begin{equation}
\label{eq2}X_\sigma (i\nu _m)=\frac 1\beta \sum_{n=-\infty }^\infty  G%
_\sigma(i\nu _m+i\omega _n)G_\sigma (i\omega _n)\, .
\end{equation}
The electron propagator $ G_\sigma (z)$ is defined for the SIAM as $%
G(z)=[z+\mu _\sigma -V^2\Gamma _\sigma (z)-\Sigma _\sigma (z)]^{-1}$,
where $%
\mu _\sigma =\mu +\sigma h-\epsilon _f$ is the effective chemical potential
and $\Gamma _\sigma (z)$ is the local element of the Green function of the
conduction electrons.  For the Hubbard model in $d=\infty$ we have
$ G_\sigma(z)=\int d\epsilon \rho(\epsilon)[z+\mu-\Sigma_\sigma(z)-
\epsilon]^{-1}$, where $\rho(\epsilon)$ is the density of states (DOS).

We can analytically continue the sums over the
Matsubara frequencies ($\omega _n=(2n+1)\pi \beta ^{-1},\nu _m=2m\pi \beta
^{-1}$) to the real frequencies and after some manipulations we obtain the
following representations at $\beta =\infty $, $n=1$ and $h=0$
\begin{mathletters}
\label{eq3}
\begin{eqnarray}
\label{eq3a}
\mbox{Re}X(\omega _{+})& =
&\int\limits_{-\infty }^0dx\rho(x)
\mbox{Re}[G(x+\omega_{+})+G(x-\omega _{+})]\, ,\\
\label{eq3b}
\mbox{Im}X(\omega _{+})&=
&-\pi sgn\omega \int\limits_0^{|\omega |}dx\rho(x)
\rho (x-|\omega |)\, ,
\end{eqnarray}
\end{mathletters}
where $\rho (\omega )=-\frac 1\pi \mbox{Im}G(\omega _{+})$, $\omega _{+}=\omega
+i0^{+}$. The self-energy can then be represented as
\begin{mathletters}
\label{eq4}
\begin{eqnarray}
\label{eq4a}
\mbox{Re}\Sigma (\omega _{+})&=&
-\frac{U^2}2\!\!\!\int\limits_{-\infty}^0
\!\!\!dx\left\{\rho (x)\mbox{Re}\left[C(x-\omega _{+})-C(x+\omega _{+})\right]
\right.\nonumber\\
\label{eq4b}
&&\left.+\frac 1\pi \mbox{Im}C(x_{+}) \mbox{Re}\left[G(x-\omega _{+})-
G(x+\omega_{+})\right]\right\}\, ,\\
\mbox{Im}\Sigma (\omega _{+})&=&U^2\int\limits_0^{|\omega |}dx\rho
(x-|\omega |)\mbox{Im}C(x_{+})\, .
\end{eqnarray}
\end{mathletters}
The two-particle correlation function $C(z):=X(z)/(1-U^2X(z)^2)$. Equations (%
\ref{eq3})-(\ref{eq4}) represent a set of nonlinear integral equations for $%
\mbox{Re}\Sigma (\omega _{+})$ and $\mbox{Im}\Sigma (\omega _{+})$. These
equations can be
solved numerically by iterations at weak coupling \cite{ls68,mm91,js95},
but the iteration procedure breaks down as $C(0)\rightarrow
\infty $ with increasing $U$. Since $X(0)<0$ the quantity $1+UX(0)$
approaches zero at intermediate coupling. The dominant contributions to $%
\Sigma (\omega )$ then come from a vicinity of the Fermi energy ($\omega
\sim 0$) where the two-particle correlation function $C(\omega)$ is sharply
peaked. We now use the Fermi-liquid assumption that only the low-frequency
behavior around the Fermi energy is decisive for the physics of interacting
electrons around the two-particle pole and replace the denominator of
$C(\omega )$ with a quadratic polynomial
\begin{equation}
\label{eq5}1+UX(\omega)\approx UX^{\prime \prime }(0)[\Delta ^2+\omega ^2-i\pi
a\omega ]
\end{equation}
where $X(0)^{\prime \prime }:=\int_{-\infty }^0dx\rho (x)\mbox{Re}G^{\prime
\prime}(x)$, $a:= \nu ^2/X^{\prime \prime }(0)$, $\Delta ^2:= (1+UX(0))/U
X^{\prime\prime}(0)$. Here $\nu $ is the DOS of the unpertubed Green function
at the Fermi energy. The parameter $\Delta $
is an energy scale measuring dominant fluctuations in the critical region of
the two-particle pole. Note that Hamann used in \cite{ha69} the same idea of
a low-frequency expansion at the denominator of a two-particle function, but
expanded $X(\omega )$ only to linear power in $\omega $.
This difference leads to
drastic changes in the critical behavior of the solution. It is also
necessary to realize that (\ref{eq5}) is valid only if Fermi-liquid theory
holds without restrictions,
i.e. there are no other relevant energies except for the Fermi one.

Inserting (\ref{eq5}) in (\ref{eq4}) we obtain in leading order of the limit
$\Delta \rightarrow 0$%
\begin{mathletters}
\label{eq6}
\begin{eqnarray}
\label{eq6a}
&&\mbox{Re}\Sigma (\omega _{+})=\mbox{Re}G(\omega)J(\infty)
-\rho(\omega)\mbox{sgn}\omega K(\omega)\nonumber\\[4pt]
&&=\frac{a\mbox{Re}G(\omega)}{2X^{\prime \prime }(0)}%
\int\limits_0^{\infty } dx\frac x{(\Delta ^2+x^2)^2+\pi^2a^2x^2}
-\frac{\rho (\omega )\mbox{sgn}\omega}{2X^{\prime \prime
}(0)}\int\limits_{0}^{|\omega |}dx\frac{\Delta ^2+x^2}{(\Delta
^2+x^2)^2+\pi ^2a^2x^2}\, ,\\
\label{eq6b}
&&\mbox{Im}\Sigma (\omega _{+})=-\rho(\omega)J(\omega)=
-\frac{\pi a\rho (\omega )}{2X^{\prime
\prime }(0)}\int\limits_0^{|\omega |}dx\frac x{(\Delta ^2+x^2)^2+\pi
^2a^2x^2}\, .
\end{eqnarray}
\end{mathletters}
We see that integral equations (\ref{eq4}) turned algebraic, where only two
positive parameters $\Delta $ and $X^{\prime \prime }(0)$ are expressed as
integrals over the products of the full Green function $G(\omega )$. The
parameter $X^{\prime \prime }(0)$ is proportional to the effective mass ($%
-\Sigma ^{\prime }(0)$) of quasiparticles from Fermi-liquid theory and
can be assumed as an effective mass of electron-hole pairs.
The energy $\Delta $ is a new relevant scale for the two-particle
scattering. Although (\ref{eq6}) is strictly valid only in the
limit $\Delta \rightarrow 0$, we can extrapolate it
also to the weak coupling, $U\rightarrow 0,\Delta \rightarrow \infty $.
Such a theory then fulfills Fermi-liquid assumtions, i.~e.
$\mbox{Im}\Sigma(\omega)\sim -\omega^2$ and $\mbox{Re}\Sigma(\omega)\sim
-\omega$ as $\omega\to 0$, and
$\mbox{Im}\Sigma(\omega)\sim -1/\omega^2$ and $\mbox{Re}\Sigma(\omega)\sim
1/\omega$ as $\omega\to \infty$. Approximation (\ref{eq6}) represents
the simplest Fermi-liquid theory with frequency dependent self-energy
determined essentially from algebraic equations. It may serve as an
alternative to the recently proposed approximations trying to clarify the
way Fermi liquid breaks down at strong coupling of the Hubbard model
\cite{gk92,zrk93,gkr93,eh90,wc94}.

The integrals in (\ref{eq6}) can be performed explicitly. To simplify
the studied equations we confine our analysis only to the SIAM.
If we use the standard approximation $\Gamma (\omega _{+})=-i\Gamma $
we can resolve (\ref{eq6}) analytically in the limit
$\Delta \rightarrow 0$. We find an explicit solution
\begin{mathletters}
\label{eq7}
\begin{eqnarray}
\label{eq7a}
\mbox{Im}\Sigma(\omega)&=&-\sqrt{\frac{V^4\Gamma^2}4 + \frac{J(\omega)}{
1+\left(\displaystyle\frac{J(\omega)\omega-\mbox{\small sgn}\omega K(\omega)
\mbox{\small Im}
\Sigma(\omega_+)}{J(\omega)V^2\Gamma-[J(\omega)+J(\infty)]\mbox{\small
Im}\Sigma(\omega_+)}\right)^2}}+\frac{V^2\Gamma}2\, ,\\[8pt]
\label{eq7b}
\mbox{Re}\Sigma(\omega)&=-&\mbox{Im}\Sigma(\omega_+)\frac{J(\infty)\omega
-\mbox{sgn}\omega K(\omega)[V^2\Gamma-\mbox{\small Im}\Sigma(\omega_+)]}{
J(\omega)V^2\Gamma-[J(\omega)+J(\infty)]\mbox{\small
Im}\Sigma(\omega_+)}\; .
\end{eqnarray}
\end{mathletters}
To close the approximation we complete these equations with definitions
of the parameters
\begin{mathletters}
\label{eq8}
\begin{eqnarray}
\label{eq8a}
X''(0)&=& \int\limits_{-\infty}^0 dx\rho(x)\mbox{Re} G''(x)
\approx \nu\mbox{Re}G'(0)\, ,\\
\label{eq8b}
\Delta^2&=&\frac 1{UX''(0)}\left[ 1+2U\int\limits_{-\infty}^0 dx\rho(x)
\mbox{Re}G(x) \right]\, .
\end{eqnarray}
\end{mathletters}
The set of equations (\ref{eq7}) and (\ref{eq8}) can be solved
numerically. Contrary to
Hamann's result we reach a critical interaction $U_c\approx 3.7/\nu$ at which
$\Delta =0$ and a pole in the electron-hole Green function appears at
the Fermi energy. The pole
leads, however, to a quite different behavior of the self-consistent solution
than in the non-self-consistent RPA.
The reason for this behavior deviating qualitatively from that found in RPA on
the one side and by Hamann on the other side
lies in a breakdown of Fermi-liquid theory around $U_{c}$.
The expansion coefficient $X^{\prime\prime}(0)$, neglected by Hamann,
diverges at the critical point. Namely
$\Delta X^{\prime \prime }(0)\rightarrow\alpha_{c}>0$ and the effective
mass of quasiparticles becomes infinite. Hence the low-frequency approximation
(\ref{eq5}) indicates a sharp transition between weak and strong coupling
regimes. The existence of a sharp transition is, however, incompatible with a
strong-coupling solution, unless the critical point
represents a metal-insulator transition, i.e. the
DOS at the Fermi energy vanishes with $\Delta\to 0$. Namely, it is easy
to show that the solution at and above $U_{c}$ does not possess
Fermi-liquid properties and
\begin{mathletters}
\label{eq9}
\begin{equation}
\label{eq9a}
\mbox{Re}\Sigma(0^-)=-\mbox{Re}\Sigma(0^+)>0\,
\end{equation}
and when $|\mbox{Re}\Sigma(0)|<w$, where $w$ is a half bandwidth, then
\begin{equation}
\label{eq9b}\mbox{Im}\Sigma(i0^+)<0\; .
\end{equation}
\end{mathletters}
The analyticity assumption for the expansion (\ref{eq5}) hence does not
hold any longer.

Analyzing the equations at $U\ge U_{c}$ we find that (\ref{eq5}) must be
replaced by
\begin{equation}
\label{eq10}
1+UX(\omega)\approx UX'(0)\left[ \Delta+|\omega|-i\pi a\omega \right]
\end{equation}
reflecting a nonanalyticity of the particle-hole bubble at low
frequencies at strong coupling. The weak-coupling and strong-coupling
ansatzes (\ref{eq5}) and (\ref{eq10}), respectively, are
evidently incompatible
and do not allow a continuous matching. There is no critical point from
the strong-coupling side as in Hamann's analysis. We must hence conclude
that the above analysis based on (\ref{eq5})
is incomplete and does not lead to the actual strong-coupling asymptotics
of self-consistent diagrammatic approximations.

Since the real part of the self-energy experiences a jump and the
imaginary part acquires a nonzero value at the critical point, expansion
(\ref{eq5}) around $\omega =0$ in this form becomes meaningless. The
frequency interval within which Fermi-liquid theory holds has shrunk to
zero at $U_{c}$. To assess the solution in the critical region,
$U\nearrow U_{c}$, more precisely we have to take into account two
contributions to the function $X(\omega_+)$. The first one is that from
(\ref{eq5}) and holds now only for $|\omega|<\Theta$. Energy $\Theta$
determines the interval within which Fermi-liquid theory holds. It can
be defined as a frequency where the real (imaginary) part of the
self-energy reaches its (first) extremum. It is essential that $\Theta\to 0$
as $\Delta\to 0$. Function $X(\omega_+)$ outside the interval
$[-\Theta,\Theta]$ must be newly approximated and one has to expand around the
points $\pm\Theta$. Such an expansion has its leading terms of type
(\ref{eq10}). In
the limit $\Delta\to 0$ the non-Fermi-liquid contributions from the
additive expansion around $\Theta\to 0$ more and more take over the
control of the critical behavior and preclude the critical point to be
reached. It means that the mechanism how the RPA pole is circumvented in
self-consistent approximations with two-particle bubbles is much more
complicated than anticipated in Hamann's low-energy analysis within
Fermi-liquid theory. Only the full treatment of self-consistent
diagrammatic approximations with the non-Fermi-liquid contributions
(\ref{eq10}) can produce the genuine strong-coupling asymptotics.
It is not yet clear whether the expected strong-coupling Kondo
asymptotics in the SIAM will be reproduced correctly in this way.
The result
depends namely on a detailed balance of the Fermi- and non-Fermi-liquid
contributions in the critical region and on the dispersion relation of
the underlying lattice. The complete analysis of the bubble-chain
approximation in the critical region $\Delta\to 0$ will be presented
elsewhere.

To conclude, we demonstrated that self-consistent diagrammatic
approximations at intermediate and strong coupling, where a pole in a
particle-hole correlation function is approached, show interesting
behavior going beyond Fermi-liquid theory. We analyzed the
self-consistent version of the bubble-chain approximation for the
Anderson and Hubbard models and proposed a simple analytic solution with
Fermi-liquid properties at weak coupling. However, to obtain the
asymptotics at strong coupling, it is necessary to restrict the validity
of Fermi liquid theory to an interval around the Fermi energy,
$\omega\in[-\Theta,\Theta]$,  vanishing when the critical point is
reached. The contributions obtained from Fermi liquid theory alone are
insufficient to suppress the two-particle singularity and to reflect the
strong-coupling behavior. Differences in solutions of (\ref{eq6})
and that of Hamann from ref. \cite{ha69} show how a delicate problem it is
to find the pertinent asymptotics of the Anderson and Hubbard models in the
critical region of a two-particle pole crucial for the transition
from the weak to the strong coupling. Neither of the above analytic
solutions can yet be seen as exact at strong coupling within the chosen
self-consistent approximations.

The work was supported in part by the grant No. 202/95/0008 of the Grant
Agency of the Czech Republic.

\end{document}